\documentstyle[aps,preprint]{revtex}

\begin{document}
\begin{center}{\Large{\bf Continuous phase transition and negative
specific heat in finite nuclei}}\\
\vskip 1.0cm
J. N. De$^{1,2}$, S. K. Samaddar$^{1}$, S. Shlomo$^{2}$ and J. B.
Natowitz$^{2}$\\
$^1$Saha Institute of Nuclear Physics, 1/AF Bidhannagar, Kolkata 700064, India\\
$^2$The Cyclotron Institute, Texas A$\&$M University, College Station,\\
Texas 77843, USA\\
\end{center}
\begin{abstract}
The liquid-gas phase transition in finite nuclei is studied in a 
heated liquid-drop model where the nuclear drop is assumed to be 
in thermodynamic equilibrium with its own evaporated nucleonic 
vapor conserving the total baryon number and isospin of the system.
It is found that in the liquid-vapor coexistence region the pressure
is not a constant on an isotherm indicating that the transition
is continuous.  At constant pressure,
the caloric curve shows some anomalies, namely, the systems studied 
exhibit negative heat capacity in a small temperature domain. 
The dependence of this specific feature on the mass and isospin of the 
nucleus, Coulomb interaction and the chosen pressure is studied.
The effects of the presence of clusters in the vapor phase on 
specific heat have also been explored.

\end{abstract}
\vskip 1.0cm
PACS Number(s): 25.70.Pq, 24.10.Pa, 64.60.Ak 
\newpage

\section{INTRODUCTION}

The possible occurrence of liquid-gas phase transition in atomic nuclei
has aroused intense interest in recent times. For macroscopic extensive
systems, phase transitions are well defined. For microscopic systems, 
nuclei for example, the presence of the surface and the
long range Coulomb interaction add complexities in defining
the liquid-gas type phase transition normally reserved for extensive 
infinite systems.
However, over the years, there has been a large build-up of 
theoretical data in different models such as 
statistical multifragmentation \cite{bon,bon1,gro},
percolation \cite{cam},
lattice-gas \cite{das,cho}, 
or a microscopic finite temperature Thomas-Fermi model \cite{de} that are
largely in consonance with the occurrence of a liquid-gas type phase
transition in finite nuclei.
Experiments \cite{poc,ell} on nucleus-nucleus collisions also 
give signatures like the critical-like behavior of the observed fragment
partitions, the nearly flat caloric curve leading to a peaked structure
in the specific heat that are compatible with the occurrence of such 
a transition in finite nuclei. 
A coherent picture 
about the characterization of its properties such as the phase diagram
or the order of the phase transition has not, however, clearly 
emerged yet.

In a microcanonical statistical multifragmentation model, with
canonical input for fragment formation probability, Bondorf
{\it et al} \cite{bon} noted some anomalous behavior in the
caloric curve in the excitation energy range 3 to 5 MeV/A
where the slope of the curve is negative. 
In a microcanonical sampling of statistical multifragmentation, Gross
\cite{gro} noted an anomalous behavior in the caloric curve leading
to negative specific heat. Such a behavior was also seen in a
microcanonical ensemble of a symmetric $A=36$ nuclear system prepared
with antisymmetrized molecular dynamics at constant pressure \cite{fur}.
Negative heat capacity in nuclear multifragmentation 
has also been observed in a canonical model \cite{das1}.
In a microcanonical framework,
Chomaz {\it et al.} \cite{cho,ago} obtained negative specific heat
from fluctuation analysis which has been widely claimed as indicators
of a first order phase transition in nuclei. 
Exploiting the standard
Clausius-Clapeyron equation for an evaporating liquid drop, Moretto
{\it et al.} \cite{mor1} find  evidence of negative specific
heat at constant pressure only when the binding energy per nucleon of the drop
increases with mass number. In the nuclear context, from the binding
energy curve, the mass number $A$ of the drop is then less than
$\sim 60$. This maximum can, however, be controlled by inhibiting the
Coulomb contribution confining the evaporated nucleus in a box \cite{mor2}.
The analyses in Refs. \cite{mor1,mor2} and the conclusions thereof have
been achieved at the cost of some simplifying assumptions, namely, the 
system has been assumed to be a one-component, the vapor phase 
is assumed to consist of monomers and the transition
is taken implicitly to be first order. For a one-component finite system, it
is straightforward to find that the Clausius-Clapeyron equation may be written 
as
\begin{eqnarray}
\frac{dP}{dT}\left (v_g -v_l) = (s_g - s_l\right ) + 
\frac{dA_l}{dT}\left (\frac{\partial \mu_l}{\partial A_l} +
\frac{\partial \mu_g}{\partial A_g}\right ),
\end{eqnarray}
where $v, s, A$ and $\mu$ refer to the specific volume, specific entropy,
mass and chemical potential for the liquid ($l$) and gas ($g$) phases,
respectively with $A_l +A_g = A_0$, the number of particles in the system.
The last term is nonzero due to the presence of the surface and is
proportional to $A_l^{-1/3}$ (assuming the surface energy of the gas
is negligible); it has been neglected in the analyses of \cite{mor1,mor2}. 
In addition, for a simplistic analysis, 
the temperature of the system is taken to be much smaller
compared to the binding energy per particle; 
this may not be typically the case
here. 

Mean-field models have often been employed to explore
liquid-gas phase transition in infinite and finite nuclear
systems. In this model, the phase transition is found to be
continuous, both for asymmetric nuclear matter \cite{bar,mul} and
also for finite nuclei \cite{sil,lee}. Though approximate,
the model serves the purpose of giving an orientation for  
understanding some important features of liquid-gas phase transition.
 It may therefore be worthwhile to undertake
a full numerical calculation to explore whether the anomalous features in 
caloric curve or in specific heat persist relaxing the constraints
imposed in Refs. \cite{mor1,mor2}. 
We make such an attempt in this paper. 
Moreover, the vapor phase may not consist of monomers only,
it may contain various clusters along with the nucleons; the
influence of clusters on caloric curve has also been explored.

\section{THE MODEL}

The model employed in the present calculation is in the framework of
mean-field theory.  The excited nucleus 
is viewed  as a charged liquid-drop composed of $N_0$ 
neutrons and $Z_0$ protons with mass number $A_0=N_0+Z_0$. 
In its journey from the liquid to the gas
phase, the depleted nucleus is taken to be in complete thermodynamic
equilibrium with its own emanated vapor so that the total number
of neutrons and protons are conserved. To keep the description
on a simpler pedestal, we first consider nucleonic vapor only. 
Besides nucleons, the vapor may contain
clusters that would alter
the equilibrium conditions which will be reflected on the
caloric curve and the resulting heat capacity. 
In the following two
subsections we present some details of the methodology
followed under these two conditions.

\subsection{Nucleonic vapor}

The framework for studying liquid-gas phase transition
for a heated nuclear liquid-drop in equilibrium with the 
nucleonic vapor has been described in some
detail in Ref.\cite{sil}. For simplicity, the mutual
Coulomb interaction between liquid and gas was ignored there;
in the present calculations this is
taken into account.
For the sake of completeness, we present here
the relevant features of the model.
The phase coexistence is 
governed by the Gibb's conditions:
\begin{eqnarray}
P_l = P_g,\nonumber\\
\mu_n^l =\mu_n^g,\\
\mu_p^l =\mu_p^g,\nonumber
\end{eqnarray}
{\it i.e}, the pressure and the chemical
potentials of neutrons and protons are the same in both the  
liquid(l) and the gas(g) phase.
This model has some resemblance to the one used 
by Lee and Mekjian \cite{lee}
where they studied the phase surface associated with liquid-gas
phase transition incorporating the role of Coulomb and surface effects.
However, they considered a fixed spherical volume at a chosen pressure and
mapped the liquid-gas coexistence region by varying the density and
proton concentration. Their results therefore do not pertain
to a particular finite nuclear system and may not correctly describe
the subtle nuances of a phase transition in a given finite system in
a mean-field theory.

The total free energy of the nuclear system (in a single phase) at
temperature $T$ is taken as
\begin{eqnarray}
F= A_0f_{nm}(\rho, X_0,T) + F_c + F_{surf},
\end{eqnarray}
where $f_{nm}(\rho,X_0,T)$ is the free energy per particle of infinite
nuclear matter at the same density $\rho$ and 
neutron-proton asymmetry $X_0$ (=$(N_0-Z_0)/A_0$) of the total
system, $F_c$ is the Coulomb free energy, and $F_{surf}$ is the temperature
and asymmetry dependent surface free energy. The free energy of infinite
nuclear matter is evaluated with the SkM$^*$ interaction in the finite
temperature Thomas-Fermi framework. The detailed expressions for the
free energies are given in the Appendix.  

In the liquid-gas coexistence
region, the free energy $F_{co}$ is
\begin{eqnarray}
F_{co}=F^l+F^g+F_c,
\end{eqnarray}
where $F^l$ and $F^g$ are the respective free energies of the liquid 
and the gas phase in the absence of the Coulomb interaction and
$F_c$ here represents the total Coulomb free energy of the system
in the mixed phase. The free energy $F^l$ is 
\begin{eqnarray}
F^l= A_lf_{nm}(\rho^l,X^l,T) + F_{surf}^l,
\end{eqnarray}
with $A_l$, $\rho^l$ and $X^l$ as the nucleon number, density and 
neutron-proton asymmetry in the liquid phase, respectively. The expression
for $F^g$ has the same form as given in Eq. (5) but here
the surface free energy  is neglected 
because of the very low density of the gas. 

Due to its thermal motion, the liquid-drop may be located
anywhere within the spherical freeze-out volume.
The Coulomb free energy $F_c$ then depends on the distance $d$ of
the center of the liquid-drop from the center of the freeze-out volume; 
this dependence is, however, very weak as will be discussed in the
next section. 
The Coulomb free energy of the liquid 
part is taken to be that of a uniformly charged sphere of radius
$R_l=r_lA_l^{1/3}$; the radius parameter $r_l$ is related to the
liquid density $\rho^l$ as $r_l=1/(\frac {4}{3}\pi\rho^l)^{1/3}$.
The  gas is taken to be uniformly
distributed in the whole spherical freeze-out volume excluding the 
volume occupied by the liquid-drop. The total Coulomb free
energy is then given by
\begin{eqnarray}
F_c&=&\frac {3}{5}e^2\left[\frac {Z_l^2}{R_l}+\frac {Z_g^2}{R}
\left(\frac{V_l+V_g}{V_g}\right)^2-\frac {Z_g^2}{R_l}\left(\frac
{V_l}{V_g}\right)^2\right]\nonumber \\
&&+\frac {3}{2}\frac {Z_ge}{R^3-R_l^3}\left[Z_le-Z_ge\frac {V_l}{V_g}
\right]\left[R^2-R_l^2-\frac {1}{3}d^2\right],
\end{eqnarray}
with $Z_l$, $Z_g$ as the proton numbers and $V_l$, $V_g$ as the
volumes in the liquid and gas phase, respectively.
The radius $R$ of the freeze-out volume $V_f$ (=$V_l+V_g$) 
can be obtained from the
densities and particle numbers in the two phases
as obtained from the thermodynamic equilibrium conditions and conservation
of neutron and proton numbers and is given by  
\begin{eqnarray}
\frac {4}{3}\pi R^3=\frac {A_l}{\rho^l} + \frac {A_g}{\rho^g}.
\end{eqnarray}
It may be pointed out that the expression (6) with $d=0$ reduces to the
same as that used in \cite{paw} for the two-step uniform density profile.

The surface free energy of the liquid part is taken as
\begin{eqnarray}
F_{surf}^l=\sigma(X^l,T)A_l^{2/3},
\end{eqnarray}
where the temperature and asymmetry dependent surface energy
coefficient $\sigma(X,T)$ is
\begin{eqnarray}
\sigma (X,T)=\left[\sigma (0,0) -a_sX^2\right]
\left[1+\frac{3}{2}\frac{T}{T_c}\right]\left[1-\frac{T}{T_c}\right]^{3/2},
\end{eqnarray}
which is obtained by considering semi-infinite nuclear matter
in equilibrium with the nucleonic vapor at the relevant temperature
and asymmetry \cite{rav}. The values of the surface energy coefficient
of semi-infinite symmetric nuclear matter in its ground state
$\sigma (0,0)$, the surface asymmetry coefficient $a_s$ and the 
critical temperature $T_c$ for the SkM$^*$ interaction are 
taken to be 17.51, 38.6 and 14.61 MeV, respectively \cite{kol}.
In our model, at a given temperature
or pressure, the number conservation and 
the thermodynamic equilibrium constraints provide a natural confining 
volume for the vapor phase in the coexistence region. 

The unknown quantities for a given liquid-drop
size $A_l$ are the number of nucleons in the gas $A_g$, 
the neutron-proton asymmetries $X^l$ and $X^g$ for
the liquid and gas phases and their respective 
densities  $\rho^l$ and  $\rho^g$. 
The quantities $A_g$ and $X^g$ are determined from
the conservation of the baryon number and the total
isospin, respectively. The three remaining unknown quantities
are determined using the conditions given by Eq.(2) 
employing Newton-Raphson method. 
With the knowledge of these quantities the free 
energy and hence all the relevant observables can be evaluated.

\subsection{Clusterized vapor}

For a given $A_l$ and with the guess values for $X^l$, $\rho^l$ and
$\rho^g$ (to be refined through iteration in order to satisfy
Eq.(2)), the volume of the gas
$V_g$ and the number of neutrons and protons in that gas are
known. With  the knowledge of the freeze-out volume
$V_g$ and the neutron and proton number, statistical 
multifragmentation model can be employed to find the multiplicities 
of the various fragments created out of the vapor phase at the
chosen temperature. We take recourse to the
grand-canonical model; in this model, 
the multiplicity $n_i$  for the $i$-th species of the 
generated fragments is given by
\begin{eqnarray}
n_i=V_g\frac {mA_i}{2\pi \hbar^2\beta} \phi_i(\beta ) exp\left [-\beta
\left (V_c^i-B_i-\mu_n N_i-\mu_pZ_i\right )\right ]
\end{eqnarray}
where $\beta $ is the inverse of temperature $T$; $m$ is the nucleon mass; 
$A_i,N_i$ and $Z_i$ are the mass, neutron and charge numbers of the
fragmenting species $i$; $B_i$'s are the binding energies of 
the generated species; $\mu$'s are the
nucleonic chemical potentials. The internal partition
function $\phi_i(\beta)$ 
for the species $i$ with $A_i>4$ is taken as
\begin{eqnarray}
\phi_i(\beta)= \int_{\varepsilon_1^i}^{\varepsilon_2^i}
d\varepsilon^* \rho_i(\varepsilon^*) e^{-\beta \varepsilon^*}.
\end{eqnarray}
Here $\varepsilon_1^i$ is the lowest excited state and $\varepsilon_2^i$
is the lowest particle-decay threshold of the $i$-th species. In
our calculations, these energy limits are taken to be the same for
all the species with $\varepsilon_1=2$ MeV and $\varepsilon_2=8$ MeV.
For the density of states $\rho_i (\varepsilon^*)$, the Bethe
level density expression
\begin{eqnarray}
\rho_i(\varepsilon^*)=\frac{6^{1/4}}{12} \frac{g_0}
{(g_0\varepsilon^*)^{5/4}} 
exp\left (2\sqrt{\frac{A_i\varepsilon^*}{10}}\right ),
\end{eqnarray}
is used with $g_0=\frac {3A_i} {4\pi^2}$. For $A_i\leq 4$,
$\phi_i(\beta )$ is taken to be unity.
The single particle Coulomb
potential $V_c^i$ is evaluated in the complementary fragment 
approximation \cite{gro1,sat} with the appropriate liquid and
vapor charge distributions. The chemical potentials so generated
may not be the same as those of the liquid phase. The chemical
and mechanical equilibrium between the liquid and the clusterized
vapor phase are obtained through an iterative procedure 
(Newton-Raphson method) by varying $X^l,\rho^l$ and $\rho^g$.

For simplicity the nuclear interaction among the fragments are neglected.
This is insignificant due to the large freeze-out volume.
The pressure of the vapor phase is taken to be that of a perfect gas
corrected for the Coulomb interaction given by 
\begin{eqnarray}
P_g=MT/V_g + \Delta P_g,
\end{eqnarray}
where $M$ is the total fragment multiplicity 
in the gas phase and $\Delta P_g$ is the
correction to the perfect gas pressure due to the Coulomb interaction.
It is given by
\begin{eqnarray}
\Delta P_g=-\frac {\partial V_c}{\partial V_g}, 
\end{eqnarray}
$V_c$ being the  Coulomb interaction energy of the system
excluding the Coulomb self-energies of the clusters in the vapor phase;
these self-energies are included in $B_i$'s.
For the evaluation of the mutual interaction part (between liquid 
and vapor) in $V_c$, the charge distribution
in the vapor phase is taken to be uniformly 
distributed in the volume $V_g$. This corresponds to 
ensemble averaged value if the clusters are randomly distributed
within $V_g$ in an event.
The treatment of the liquid phase is the same as that described in
the previous subsection.

The excitation energy of the system is evaluated from the energy
balance condition 
\begin{eqnarray}
-B+E^*=\frac {3}{2}MT -\sum n_i B_i+
\sum n_i \langle E_i^*\rangle +E_l +V_c.
\end{eqnarray}
Here $B$ is the ground state binding energy of the system, 
$\langle E_i^* \rangle$ is the average excitation energy 
of the $i$th species 
and $E_l$ is the energy
of the liquid-drop (excluding the Coulomb energy). 
The free energy of the vapor phase is the sum of the free energies
of the fragments generated in this phase, including the contribution
due to their thermal motion. The detailed expressions are given 
in Ref.\cite{bon}.
The expression corresponding to Eq.(15) for the case of monomeric
vapor is obtained by dropping off the second and third terms on the 
right hand side and replacing $M$ by $A_g$, the number of nucleons in
the vapor. The energy of the excited liquid drop $E_l$ is the sum
of the contributions from the volume, Coulomb and surface terms. 
The expression for the volume  term is given
in the Appendix. The Coulomb energy is the same as the Coulomb free energy
as it has no explicit temperature dependence.
The surface energy $E_{surf}^l$ is given by
\begin{eqnarray}
E_{surf}^l=F_{surf}^l-T\left (\frac {\partial F_{surf}^l}
{\partial T}\right )_V.
\end{eqnarray}

\section{RESULTS AND DISCUSSIONS}

 In order to explore the characteristics of phase transition in finite
nuclei, we have chosen three representative systems, namely, 
$^{40}$Ca, $^{150}$Re and $^{150}$Nd. The symmetric systems
$^{40}$Ca and $^{150}$Re are chosen to study the mass dependence whereas 
the isobaric pairs $^{150}$Re and $^{150}$Nd offer the means to study the
effects of asymmetry on the properties of the phase transition.
To discern the effect of the long range Coulomb interaction, calculations
are also done both with and without this interaction.

The distance $d$ of the center of the liquid-drop from that of the
freeze-out volume may vary from 0 to some maximum value $d_{max}$.
The latter corresponds to the surface of the drop touching the
boundary of the freeze-out volume. The reduction in $F_c$ as $d$
is increased from 0 to $d_{max}$ is very small; it is only a few percent, 
typically less than 4\% of $F_c$. 
The results calculated with these two extreme values of $d$
are practically indistinguishable, so we report calculations
taking  $d$=0. 

In the following two subsections we present the results of our 
calculations, first for the nucleonic vapor and then 
the clusterized vapor.

\subsection{Nucleonic vapor}

 In Fig.1 (top panel), the isotherm for the symmetric nucleus 
 $^{150}$Re at $T$=7 MeV is 
displayed. The full line refers to the results with Coulomb included, 
the dashed line corresponds to those without the Coulomb term. 
The lines (full or dashed) are obtained by exploiting the thermodynamic
equilibrium conditions with the constraints of baryon number and isospin
conservation and thus correspond to the fully physical region.
The high density side where the pressure rises very sharply with density
(beyond the point C) is the fully liquid phase. The wing from B to
A and to further lower densities corresponds to the fully
gas phase. The region from B to C is the region of liquid-gas
phase coexistence. Symmetric nuclear matter behaves like a one-component
system; the phase transition there occurs at a constant pressure.
Without the Coulomb interaction even a symmetric finite nucleus
(like $^{150}$Re) does not behave like a one-component system because of
the presence of the surface; the pressure changes, 
though weakly, along the phase 
transition. At constant pressure the transition then occurs over a finite
temperature interval and the transition is thus $continuous$. 
With the introduction
of Coulomb this effect is more pronounced. The isotherms 
obtained here display a Van der Waals type of loop;
this kind of behavior has been observed earlier in an 
exact calculation for finite systems by Katsura \cite{kat} and
Hill \cite{hil}. As mentioned in Refs. \cite{kat,hil}, the loop
results from the inter-facial (surface) effects between the two phases.
The isotherms for the same system ($^{150}$Re) with Coulomb interaction
switched off are displayed in the lower panel of the figure in a
narrower ($P,\rho$) interval at two nearby temperatures. One finds that
for this system 
the pressure decreases with increasing
density in the coexistence region.
At constant pressure this  may lead to negative heat capacity
$c_p$.
At a fixed pressure,say, $P\sim 0.015$ MeV fm$^{-3}$, as the temperature of
the system is increased from 7 MeV to 7.2 MeV, the density changes
from that at the point $a$ to a higher density at the point $b$ 
leading to negative isobaric volume expansion coefficient
$\alpha =\frac{1}{V}\left (\frac {\partial V}{\partial T}\right )_P$.
Since $c_p=c_v+\alpha VT \left (\frac {\partial P}{\partial T}\right )_V$
and $\left ( \frac {\partial P}{\partial T}\right )_V$ is positive
as seen in the figure, $c_p<c_v$ and under suitable conditions (as
met in our calculations), it can be negative. 

The isotherms for the
asymmetric system $^{150}$Nd at $T$=7 MeV with and without
the inclusion of the Coulomb interaction are shown in Fig.2.
Unlike the symmetric system $^{150}$Re, here with the increase
in density, the pressure initially decreases and then increases
in the coexistence region. It appears that this difference in
the behavior of the isotherms is a reflection of the asymmetry
effect. To confirm this, we have done calculations for two
isotopes of Ca, $^{40}$Ca and $^{50}$Ca, at $T$= 7 MeV
with Coulomb interaction. The asymmetry
effect is apparent as is displayed in Fig.3. 
For clarity, the looping near the onset of the vapor phase
is shown magnified in the inset in Fig.2. As the temperature is increased,
the phase coexistence region shrinks and the points B and C come 
closer and at a certain temperature (the critical temperature
$T_c$) they merge. The magnitudes of the critical 
parameters, namely, the temperature, pressure and
density for the system $^{150}Nd$ are 13.1, 0.174 and 0.050
(all in MeV-fm units), respectively.

The variation of the liquid proton fraction $Y^l$
(=$Z_l/A_l$) at constant pressure is shown as a function of the
mass number $A_l$ of the depleting drop in the coexistence phase
in the upper two panels of Fig.4 for the systems $^{150}$Re and
$^{150}$Nd. The calculations are representative, done with Coulomb interaction
on at a constant pressure $P$=0.0014 MeV fm$^{-3}$.
In all the panels  in this figure, the dashed lines refer to results
with the monomeric vapor phase and the full lines correspond
to those with the clusterized vapor phase. The variation of $Y^l$
with $A_l$ in the two systems are somewhat different. The nucleus
$^{150}$Re being highly proton-rich initially sheds off more protons
to sustain chemical equilibrium with reduction in $Y^l$ while for
the more neutron-rich nucleus $^{150}$Nd, the proton fraction
in the liquid-drop increases monotonically with its reduction in size.
These results are seen to be insensitive to the particular choice
of the vapor phase.

In order to maintain chemical equilibrium in the two phases,
the neutron and proton chemical potentials along the coexistence line
behave in a relatively complex fashion; they are displayed in the
bottom panels of Fig.4 as a function of the proton fraction in
the respective phases. The vertical arrows on the abscissa mark 
the proton fraction of the total system. The direction of the arrow along a 
curve signifies depletion of the liquid phase, {\it i.e}, the direction of 
increasing excitation energy. The
thick lines represent the results for protons and the thin lines 
are those for the neutrons. The results for the liquid and gas
phases are well demarcated and are labelled in the figure. The
proton fraction in the two phases are quite different indicating
isospin distillation \cite{xu} in the coexistence phase. This is more
prominent in the asymmetric system ($^{150}$Nd).
Contrary to the neutron-rich
gas phase for the asymmetric system,  the symmetric 
nucleus $^{150}$Re has a proton-rich gas phase. The results
again are seen to be almost independent of the choice of the 
constituents of the vapor phase we have taken.

The calculated caloric curves at constant pressure for the 
symmetric systems $^{40}$Ca and $^{150}$Re are displayed
in Fig.5 with the Coulomb interaction switched on and off. The upper
panel corresponds to a pressure $P= 0.0014$ MeV fm$^{-3}$ and the
middle panel is for a pressure twenty times higher, {\it i.e.}, 
at $P=0.028$ MeV fm$^{-3}$ for the lighter system $^{40}$Ca. 
The caloric curves show some anomalous
features. At excitation energies lower than the one marked by 
point A, the system is in the fully liquid phase; here the rise in
excitation with temperature is like that of a Fermi-gas. The region A to
B corresponds to the liquid-vapor coexistence. As the system moves from
A to B, the nucleus increasingly gets depleted in size with 
emanation of vapor and beyond B, it is in the fully vaporized 
state.  The caloric curves for the heavier system
$^{150}$Re with the Coulomb interaction switched on and off are shown
in the bottom panel at a pressure $P$=0.0014 MeV fm$^{-3}$. The
characteristics of these caloric curves are very similar to those
for the lighter system $^{40}$Ca. For both the lighter and the
heavier symmetric nuclei we consider, it is seen that the temperature 
decreases with the excitation energy over the whole coexistence
phase; this would lead to negative heat capacity. It is further noted 
that the negative slope of the caloric curve is amplified
with the inclusion of the Coulomb interaction.

Fig.6 displays the caloric curves for the asymmetric system $^{150}$Nd
at $P=0.0014$ MeV fm$^{-3}$ with and without the inclusion
of the Coulomb interaction. The presence of a Van der Waals type of
loop in the isotherms for finite systems 
for fully physical results obtained with consideration of complete 
thermodynamic equilibrium between the liquid and the gas phase has 
been mentioned before. The presence of this loop
is expected to produce negative slope in the caloric curve.
To elaborate this further, calculations have been performed  by switching
off both the Coulomb ($C$) and the surface ($S$) effects. This
actually corresponds to the case of infinite asymmetric nuclear
matter. The dashed line represents the results considering 
single-phase ($sp$) over the whole excitation energy domain considered.
Here the temperature passes through a maximum and a minimum very
similar to the variation of pressure with density at constant
temperature, the slope of the caloric curve being negative
in between the maximum and the minimum. 
 For an infinite system the loop in the isotherm
is absent with appropriate consideration of the liquid-vapor
coexistence phase ($cp$) and the negative slope in the caloric
curve is then expected to be absent here. This is really the case as
can be seen from the caloric curve shown by the full line
where the excitation energy domain 
between the points $A$ and $B$ refer to liquid-vapor
coexistence region. 
The caloric curves with successive inclusion
of the surface and the Coulomb effect are shown by the open circles
and the filled circles, respectively. 
The characteristic features of these
caloric curves are grossly the same as those of the symmetric systems
$^{40}$Ca and $^{150}$Re discussed before. However, one
distinct difference is noticeable.  Unlike the
symmetric systems $^{40}$Ca or $^{150}$Re where with Coulomb 
on or off, the
negative heat capacity extends over the whole coexistence phase,
for the asymmetric system $^{150}$Nd, this occurs for $A_l<30$
with Coulomb off
and with inclusion of Coulomb, the corresponding $A_l$ is around 100.

The results on caloric curve presented above show that both
the Coulomb interaction and the asymmetry play important role in
determining its detailed characteristics. The asymmetry tends
to produce caloric curve with a positive slope. 
This is evident from the observation
that for an asymmetric system like $^{150}$Nd, the negative slope
occurs only after some nucleons (predominantly neutrons)
are evaporated and the residual nucleus is closer to a
symmetric one whereas for a symmetric system like $^{40}$Ca
or $^{150}$Re, the slope is negative throughout the coexistence phase. 
The Coulomb interaction, on the other hand,
tends to enhance the negative slope as is clear from the
results presented in Figs.5 and 6. The asymmetry of the evaporated
gas is enhanced with the asymmetry of the nucleus whereas it is
reduced by the Coulomb interaction. The delicate dependence
of the temperature and excitation energy on the asymmetry
in the two phases to maintain phase coexistence seems to
be responsible for the features of  caloric curve as
mentioned above.

The specific heat at constant pressure $c_p 
\left (=\frac {1}{A}\frac {d}{dT}[E+PV]_P\right)$ for the systems $^{40}$Ca
and $^{150}$Nd at $P=0.0014$ MeV fm$^{-3}$ are displayed in Fig.7 and
Fig.8 as a function of excitation energy per particle. The upper panels 
correspond to results with Coulomb off, the lower panels the same
with Coulomb on. There are two discontinuities in the specific heat, one
at lower excitation and the other at higher excitation (marked by arrows)
and the specific heat is negative within this range. These 
discontinuities refer to the change in the sign of the slopes of the
caloric curves. The discontinuity at the higher excitation always occurs
at the point of complete vaporization of the system, the one at the lower
excitation depends on the system (symmetric or asymmetric) and on the
choice of the interaction (Coulomb on or off). 
The negative specific heat occurs in a 
small temperature interval, typically $\sim$ 1.0 to 1.5 MeV
for the cases we have considered.

Analysis of fluctuations of kinetic energy of the fragments
\cite{ago} and the direct measurement of the caloric
curve as well as the analyses of a number of other 
observables by TAMU group 
\cite{ma} indicate that the heat capacity is negative only in a 
small excitation energy domain, around 
3 to 7 MeV/A in Ref.\cite{ago} and around 5 to 6 MeV/A in
Ref.\cite{ma}.
Our calculated results are in contradiction with this finding
as the calculated specific heat at constant pressure
remains negative up to excitation energy as high as
14 MeV/A. This may be traced back partly to the inadequacy
of the mean-field model we adopt that masks the fluctuations.
Furthermore, in our model the vapor phase consists of monomers only. 
In reality nucleons are emitted along with various fragments.
This would modify the equilibrium conditions like pressure,
chemical potentials and
temperature for a given excitation energy of the fragmenting system
and in turn the caloric curve and the associated heat capacity
would be affected. This aspect has been 
dealt with in the next subsection. 
The actual conditions prevailing
in a fragmentation scenario may neither be isobaric nor isochoric.
This may also modify the behavior of the specific heat
appreciably leaving some room for uncertainties
in the comparison of the existing theoretical results
with the experimental findings.

 In Fig.9, the calculated entropy per particle $S/A$ 
at constant pressure $P= 0.0014$ $MeV$ $fm^{-3}$ is shown as
a function of temperature for the system $^{150}$Nd. The notations
used are the same as those in Fig.6. The results for the single-phase
calculation (dashed-line) for infinite asymmetric matter having the
neutron-proton asymmetry same as that of $^{150}$Nd have some
anomalous behavior. In a large domain of temperature it is seen
that the entropy increases with the decrease of temperature. This
unphysical character vanishes with the appropriate inclusion
of the liquid-vapor coexistence phase as shown by the full line,
the region $A$ to $B$ being 
the region  of coexistence for the two phases. With the inclusion of the
surface and/or the Coulomb effect, the        
entropy-temperature curve for $^{150}$Nd displays a back-bending 
(negative slope) even with the inclusion of the
coexistence phase as shown by the open circles and the filled circles.
The change in entropy does not show any discontinuity with 
temperature though there are marked changes in slope; 
the phase transition is then continuous.
The temperature window for the persistence of the  negative slope
of the entropy curve is the same as that seen in  the caloric curve.
The dependence of entropy on the excitation energy at constant
pressure is shown
in Fig.10 for the different cases studied as mentioned
in the context of Fig.9. It is
seen that for all the cases the entropy increases monotonically
with the excitation energy as expected.

Before closing this subsection, it may be worth mentioning that
liquid-gas coexistence may occur with a bubble-configuration
where the gas is enclosed in a shell of liquid. 
In nuclear matter, the minimum size for a possible bubble to
occur is similar to the size of a heavy nucleus \cite{shl}.
For finite systems, our calculation shows that the drop-configuration is 
favored over the bubble. The
free energy for the bubble-configuration is found to be much higher 
compared to that of the corresponding drop-configuration
with the same $A_l$. This is due to the large surface free energy
in the bubble-configuration with two liquid surfaces having large
radii.

\subsection{Clusterized vapor}

For the clusterized vapor, the results for 
the caloric curve are displayed for
the symmetric system $^{150}$Re (top panel) and that for the
asymmetric system $^{150}$Nd (bottom panel)
at $P$=0.0014 MeV fm$^{-3}$ in Fig.11.
To discern the effect of clusters, the caloric curves for
nucleonic vapor are also presented (filled circles) along with
those for the clusterized vapor (filled triangles). As in Fig.5,
the region $A$ to $B$ corresponds to the liquid-vapor coexistence
region. The chemical potential profiles with proton fraction
in the coexistence phase do not differ much for the two choices of
the vapor phase as is seen in Fig.4. 
The basic features of the caloric
curve are also not altered with the inclusion of clusters. Thus
all the remarks for caloric curve made in the previous
subsection are also valid here. One important difference, however,
lies in the occurrence of complete vaporization of the system at a
relatively lower excitation energy with consideration of
clusters. This is self-evident as the clusters are bound systems. 
With increasing excitation energy 
clusters dissolve into nucleons and 
the caloric curves tend to merge.

In Fig.12 the heat capacities at constant pressure ($c_p$) are
displayed corresponding to the caloric curves shown in Fig.11.
The arrows indicate the positions of excitation energies
corresponding to the change in sign of $c_p$ and between
the arrows it is negative. The upper bound of the excitation energy
for $c_p$ being negative is reduced significantly (from $\sim$
14 MeV to 10 MeV) with the inclusion of clusters for reasons
already stated in connection with caloric curve. However, this
reduction is not sufficient to match the experimental finding.

The entropy for $^{150}$Nd as a function of 
temperature is shown in the top
panel of Fig.13. As in the case of monomeric vapor, the
back bending, {\it i.e}, increase of entropy with reduction 
in temperature, persists 
in a narrow temperature interval 
even after inclusion of fragments.
The bottom panel shows
the entropy as a function of excitation energy for the same
system. With clusterized vapor, the entropy is a little larger
compared to monomeric vapor showing that the clusterized configuration
is more favorable.

\section{CONCLUDING REMARKS}

Within a mean-field framework, we have studied the liquid-gas phase
transition in finite nuclei with exact conservation of baryon number
and isospin. As in asymmetric nuclear matter, since at constant pressure, 
the transition occurs over a finite temperature domain and as
the entropy shows no discontinuity with temperature, we
conclude that in the model studied, the said transition in finite
nuclei is continuous. However, unlike 
bulk systems, a Van der Waals type loop in the isotherm
is observed in the coexistence region because of the finite
size effects.
 This loop, arising from the
thermodynamic equilibrium conditions results in negative specific
heat at constant pressure in a small temperature domain.
Moretto {\it et al.} \cite{mor2} find negative specific
heat only for nuclei with $A\leq $60; we find distinct evidence
of negative specific heat for the considerably heavier systems. 
In Ref.\cite{mor2}, the size of the system refers to the residual
evaporated drop, this may be identified with the depleted drop of size
$A_l$ in our calculations rather than the total mass $A_0$ of the nucleus.
Aside from this fact, the main reason for the above discrepancy   
lies in the exact computation of the equilibrium configurations
in the liquid and gas phase in our calculations.
The presence of  back bending in the caloric
curve or a negative heat capacity in microcanonical or 
canonical formulations has been taken as a tacit evidence that 
the liquid-vapor phase transition in finite nuclei is first order.
In fact, there is evidence of a first order phase transition
in two-component systems \cite{das2} in multifragmentation
calculations based on the statistical model. In the mean-field model,
however, the fluctuations implied in the fragmentation calculations
are absent, the phase transition is seen to be continuous and
a negative specific heat for finite nuclei is not incompatible with it.
The qualitative features of the results remain the
same if instead of pure nucleonic vapor, vapor with clusters is
considered. The important change with inclusion of cluster is
the significant reduction in the maximum excitation energy 
(from around 14 to 10 MeV/A) up to which the heat capacity
is negative. This maximum excitation is
still higher compared to the 
experimental finding ($\sim $ 7 MeV/A); aside from the inadequacy of
the mean-field model,   
a possible reason may be due to  the
condition under which fragmentation occurs that may neither be isobaric
nor isochoric.

\vskip 0.7cm
\begin{center}
{\bf ACKNOWLEDGMENTS}
\end{center}
J.N.D. gratefully acknowledges the kind hospitality at the Cyclotron Institute
at Texas A $\&$ M University where the work was partially done. J.N.D. and
S.K.S.  acknowledge the Department of Science \& Technology and
the Council of Scientific and Industrial Research
of the Government of India, respectively, for the financial support.
This work was supported in part by the U.S. National Science Foundation 
under Grant No. PHY-0355200, the U.S. Department of energy
under Grant No. DE-FG03- 93ER40773 and by the Robert A. Welch Foundation.

\vskip 0.7cm

\appendix
\def\singleappendix#1{\section*{#1}\renewcommand{\theequation}{A\arabic{equation}}}
\singleappendix{Evaluation of free energy}

The different components occurring in the total free energy $F$
in the single phase given in (Eq.3) are the volume, Coulomb and surface terms.
The explicit expressions for them are as follows.

\begin{enumerate}
\renewcommand{\theenumi}{(\roman{enumi})}
\item {\bf The volume term:}   The free energy per particle
of infinite asymmetric nuclear matter $f_{nm}(\rho ,X_0, T)$ is
\begin{eqnarray}
f_{nm}=e_{nm}-Ts_{nm},
\end{eqnarray}
where $e_{nm}$ and $s_{nm}$ are the energy and entropy per particle.
The energy $e_{nm}$ is given by
\begin{eqnarray}
e_{nm}=\varepsilon_{nm}/\rho,
\end{eqnarray}
where $\varepsilon_{nm}$ is the energy density,
\begin{eqnarray}
\varepsilon_{nm}=\sum_{q=n,p} \frac {\hbar^2}{2m_q}\tau_q+\varepsilon_i,
\end{eqnarray}
where $m_q$ is the nucleon mass and $n$ or $p$ stand for neutron or proton.
Here $\varepsilon_i$ is the interaction energy density.
The terms under the summation
represent the kinetic energy density expressed as
\begin{eqnarray}
\tau_q=\frac {2m_q}{\hbar^2}A_{T,q}TJ_{3/2}(\eta_q).
\end{eqnarray}
The fugacity $\eta_q$ is related to the nucleon density  as
\begin{eqnarray}
\rho_q=A_{T,q}J_{1/2}(\eta_q),
\end{eqnarray}
with
\begin{eqnarray}
A_{T,q}=\frac {1}{2\pi^2}\left (\frac {2m_qT}{\hbar^2}\right )^{3/2}, 
\end{eqnarray}
and $J$'s are the Fermi integrals given by
\begin{eqnarray}
J_k(\eta)=\int_0^\infty \frac {x^k}{1+e^{(x-\eta)}}dx.
\end{eqnarray}
For the SkM$^*$ interaction, the interaction energy density for nuclear matter
is  \cite{bra}
\begin{eqnarray}
\varepsilon_i&=&\frac {1}{2}t_0\left [(1+\frac {1}{2}x_0)\rho^2
-(x_0+\frac {1}{2})(\rho_n^2+\rho_p^2)\right ]
+\frac {1}{12}t_3\rho^\alpha \left [\rho^2-\frac {1}{2}(\rho_n^2
+\rho_p^2)\right ]\nonumber \\
&&+\frac {1}{4}\left [t_1+t_2\right ]\tau \rho
+\frac {1}{8}\left [t_2-t_1\right]\left (\tau_n\rho_n+\tau_p\rho_p\right),
\end{eqnarray}
with $\tau =\tau_n+\tau_p$ and $\rho =\rho_n+\rho_p$. The values of the 
parameters in Eq.(A8) are given in Ref.\cite{bra}.
The entropy per nucleon $s_{nm}$ is
\begin{eqnarray}
s_{nm}=\frac {1}{\rho}\sum_q\left [\frac {5}{3}A_{T,q}J_{3/2}(\eta_q)
-\eta_q\rho_q\right ].
\end{eqnarray}

\item {\bf The Coulomb term:}  The Coulomb free energy is taken to
be that of a uniformly charged sphere,
\begin{eqnarray}
F_c= \frac {3}{5}Z_0^2e^2/R,
\end{eqnarray}
where $Z_0e$ is the total charge of the system with radius $R$.

\item {\bf The surface term:}  The surface free energy $F_{surf}$
is given by
\begin{eqnarray}
F_{surf}=\sigma(X_0,T)A_0^{2/3},
\end{eqnarray}
where the expression for the surface energy coefficient $\sigma (X,T)$
is given by Eq.(9).
\end{enumerate}

For the vapor phase, the expressions for volume and coulomb energy 
have the same form; the surface free energy is taken to be zero
because of its very low density.

\newpage
\centerline 
{\bf Figure Captions}
\begin{itemize}
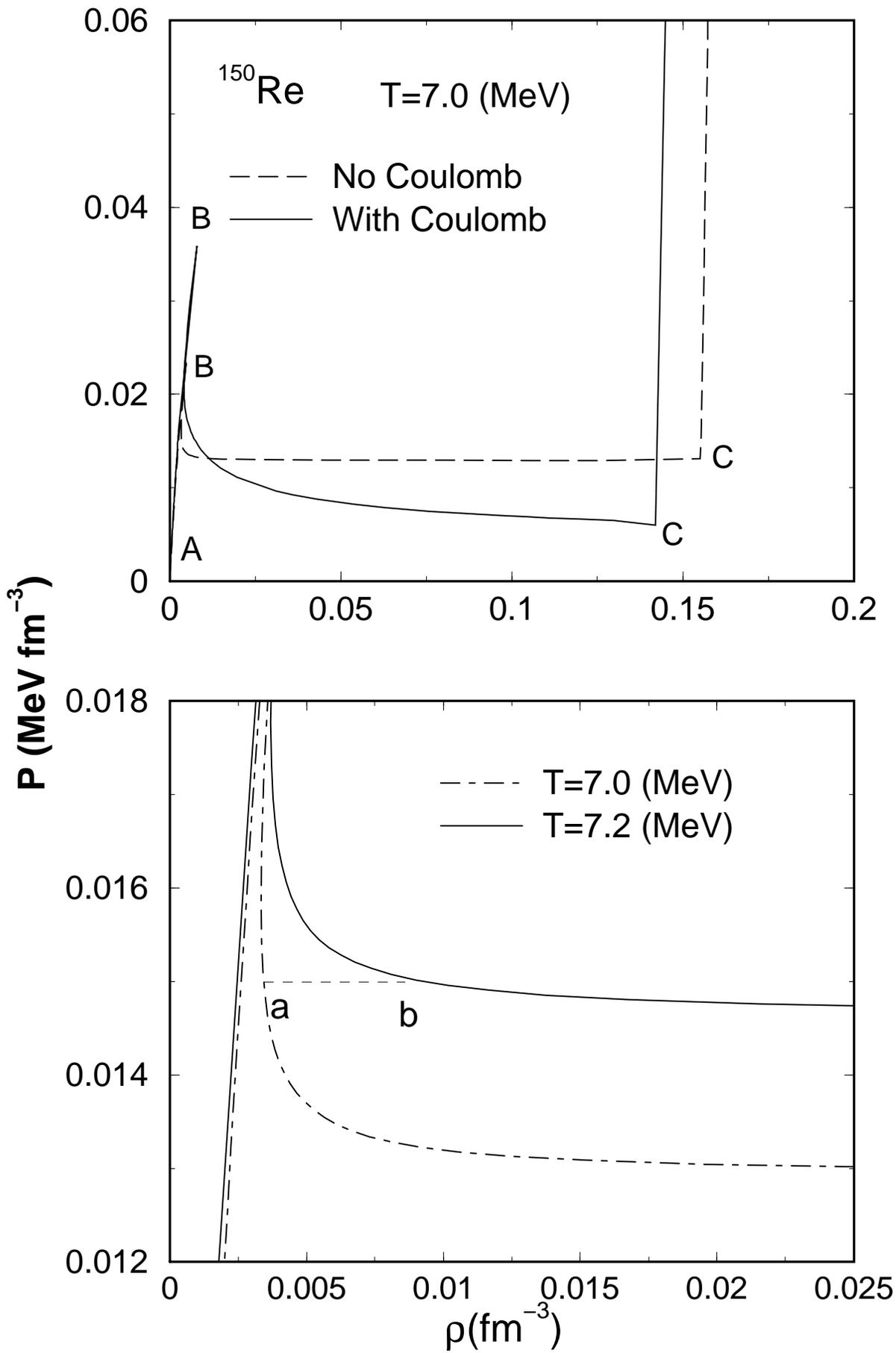
\item[Fig.\ 1] The isotherms for the system $^{150}$Re at $T=$ 7.0 MeV 
with and without the Coulomb interaction (top panel). The
bottom panel shows the same at $T=$ 7.0 and 7.2 MeV with Coulomb off
in a narrow density interval. For the horizontal line $a-b$, see text. 
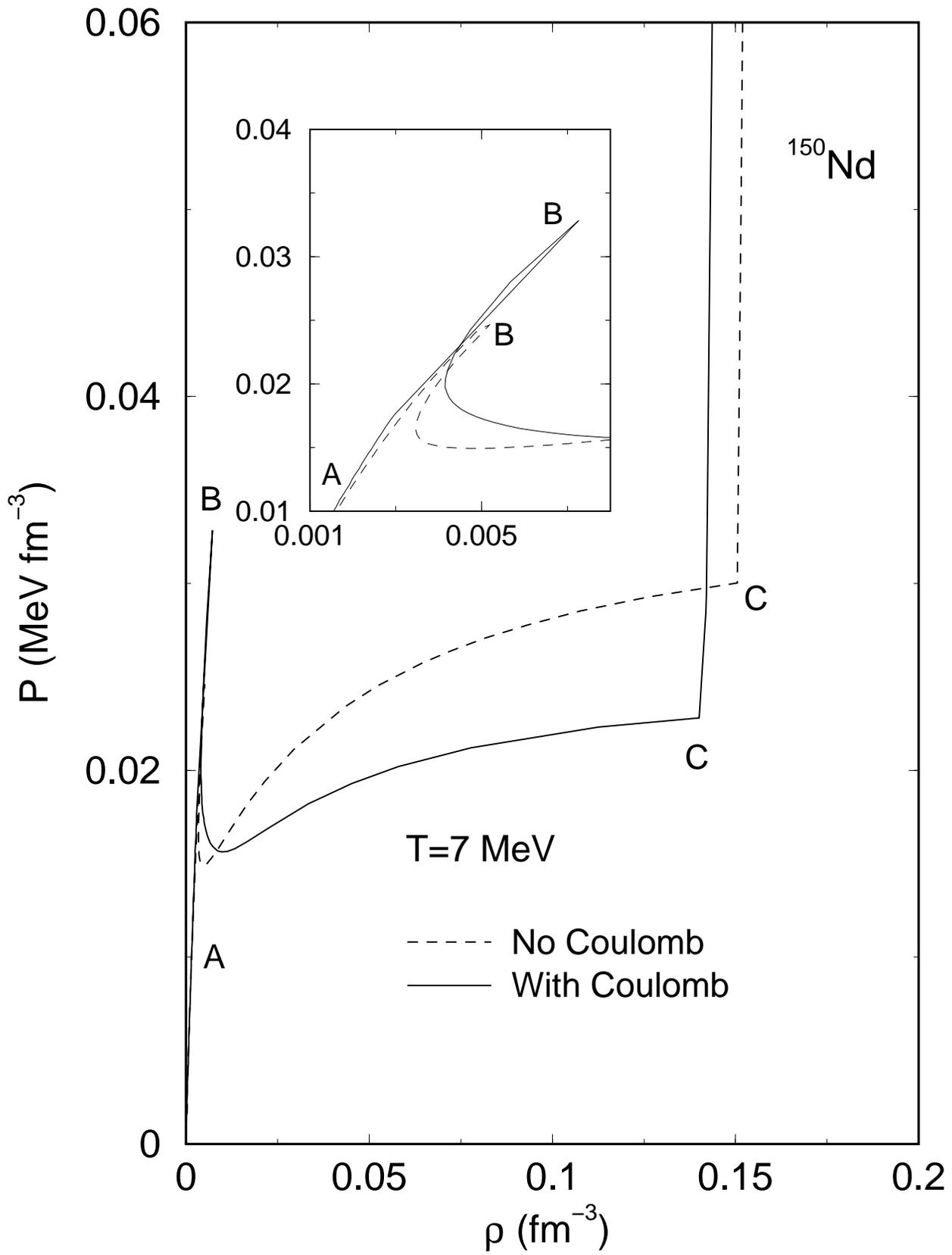
\item[Fig.\ 2] The isotherms for the system $^{150}$Nd with and
without Coulomb interaction. The inset magnifies the loop           
in a narrow density region.
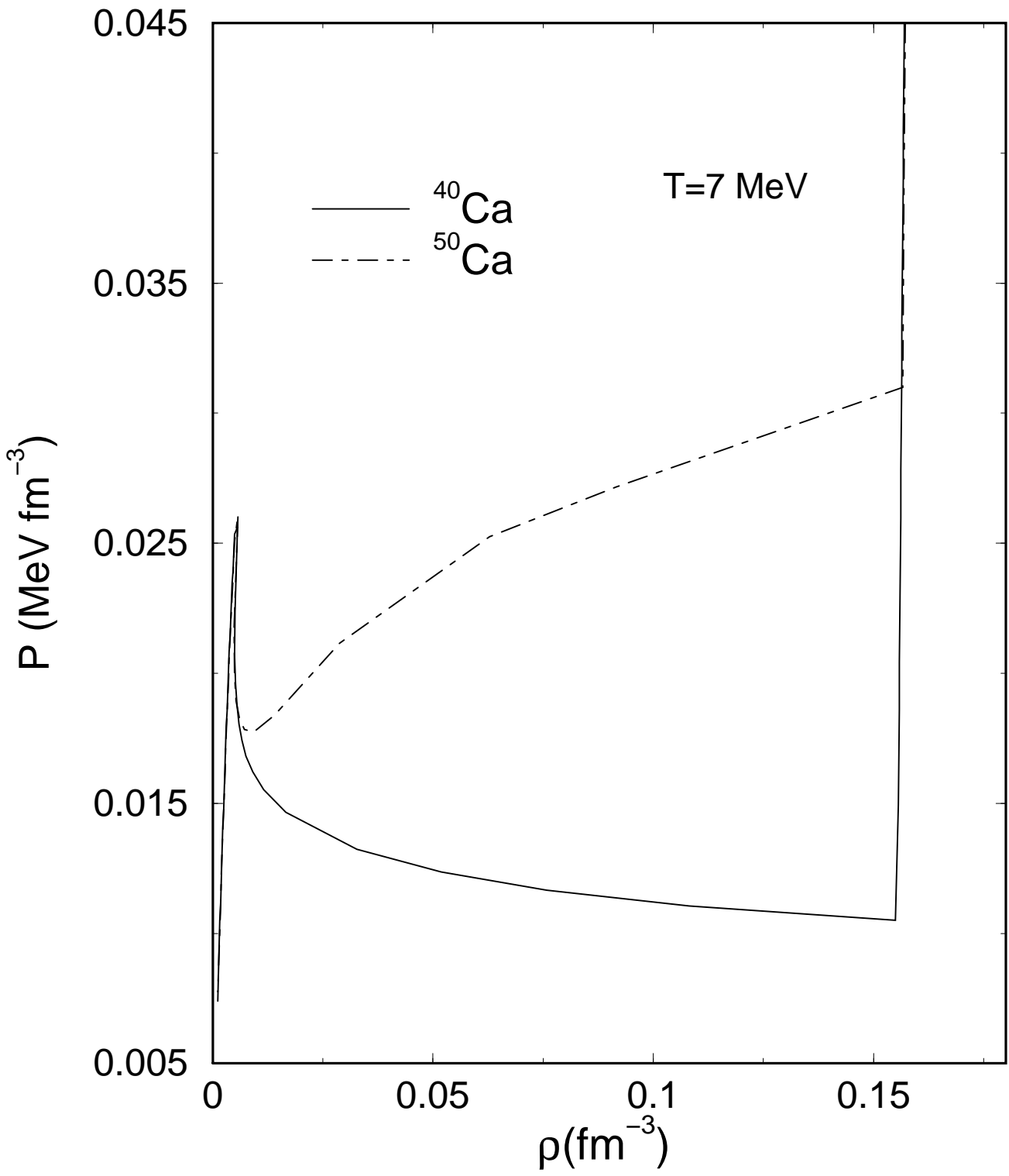
\item[Fig.\ 3] The isotherms for the nuclei $^{40}$Ca and $^{50}$Ca
at T=7 MeV with Coulomb interaction. 
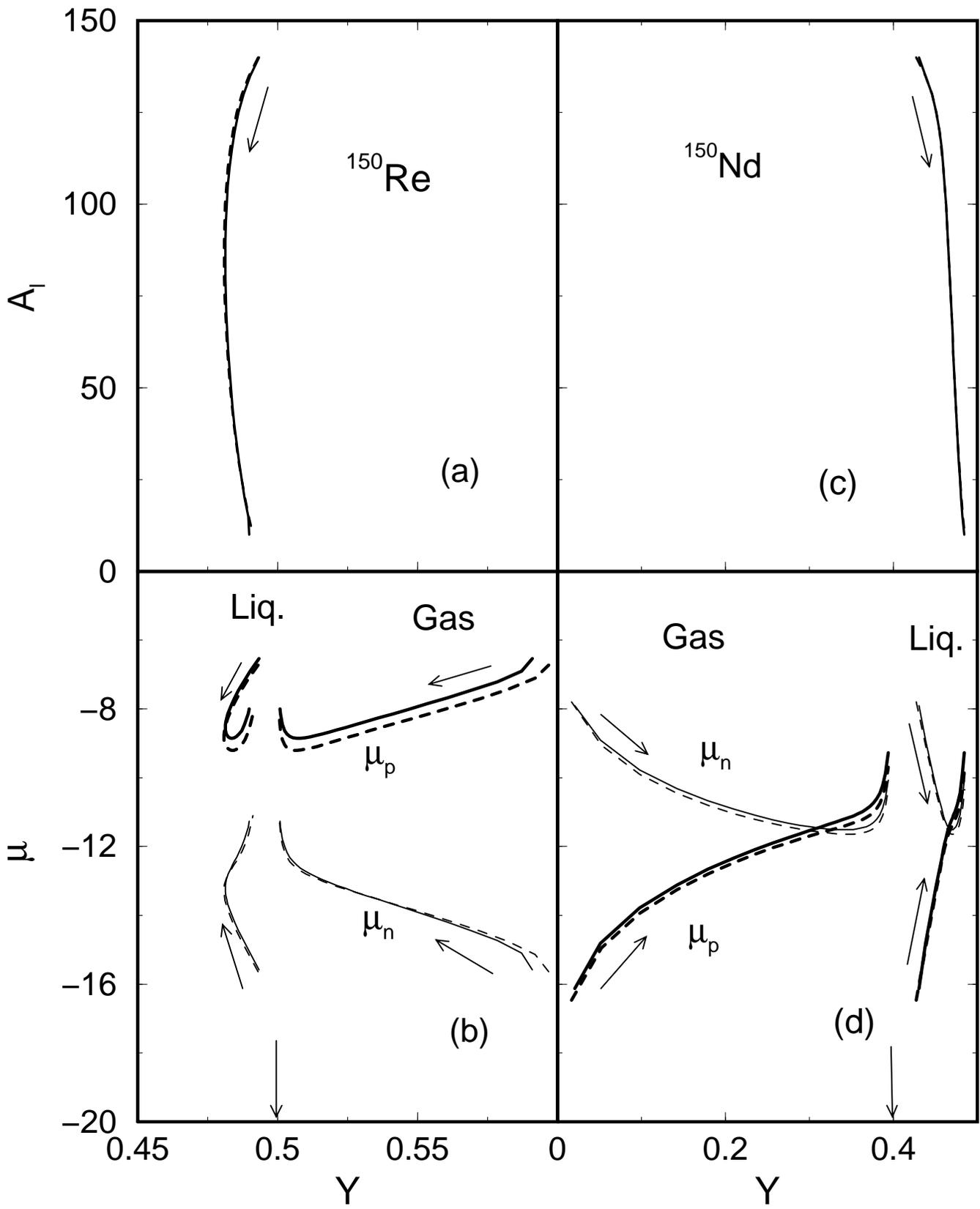
\item[Fig.\ 4] The upper panels display the proton fraction $Y^l$
of the liquid-drop as a function of its depleting mass number $A_l$.
The lower panels show the neutron and proton chemical potentials 
in the liquid and gas phase as a function of the proton fraction
in the respective phases. The liquid and gas phase results are 
well separated and are marked in the figure. The vertical arrows
on the abscissa show the proton fraction of the total system. The 
results correspond to constant pressure $P$= 0.0014 MeV fm$^{-3}$.
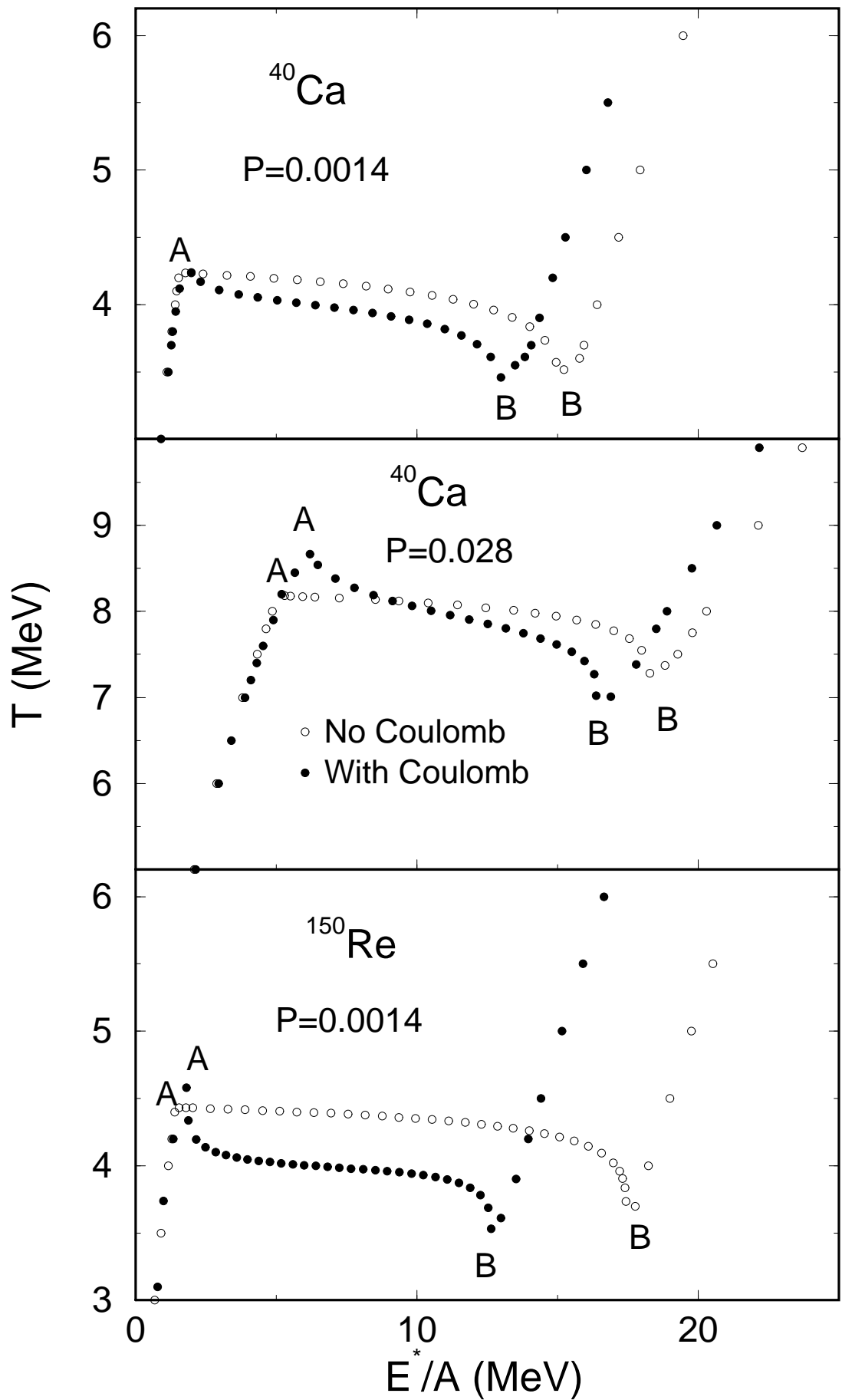
\item[Fig.\ 5] The caloric curve for the system $^{40}$Ca at $P=0.0014$
 MeV fm$^{-3}$ with (filled circles) and without (open circles) Coulomb
interaction (top panel). The middle panel displays the same at $P=0.028$
 MeV fm$^{-3}$. The bottom panel is the same as in the top panel, but
for the system $^{150}$Re. 
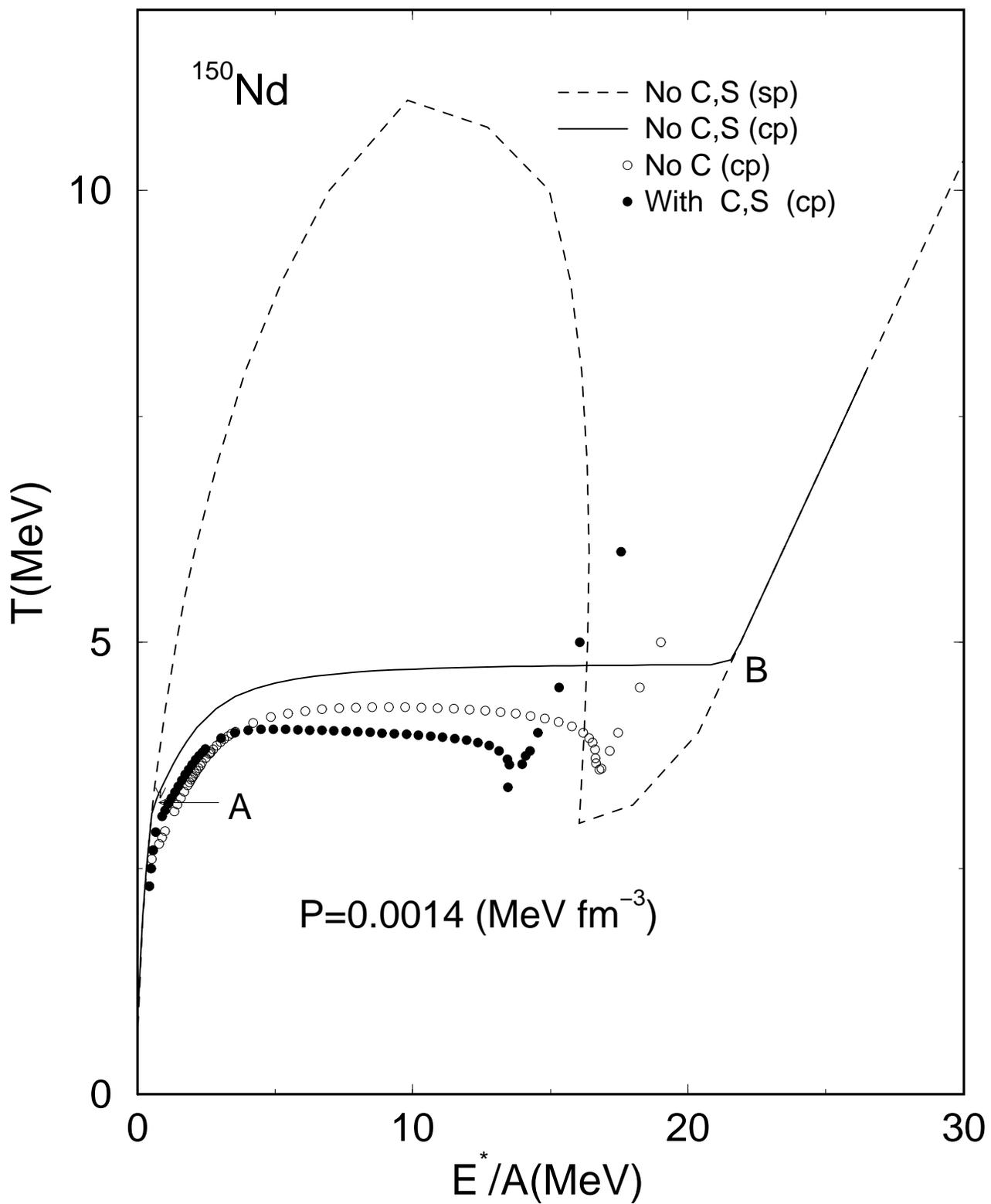
\item[Fig.\ 6] The caloric curves for the system $^{150}$Nd at
$P$=0.0014 MeV fm$^{-3}$. The dashed line is obtained after switching off
the Coulomb and the surface (C,S) effects in the single phase (sp),
the full line refers to that with the
coexistence phase (cp), the open circles 
correspond to the one after switching on the surface
effect and the filled circles 
refer to caloric curve with both Coulomb and surface on.
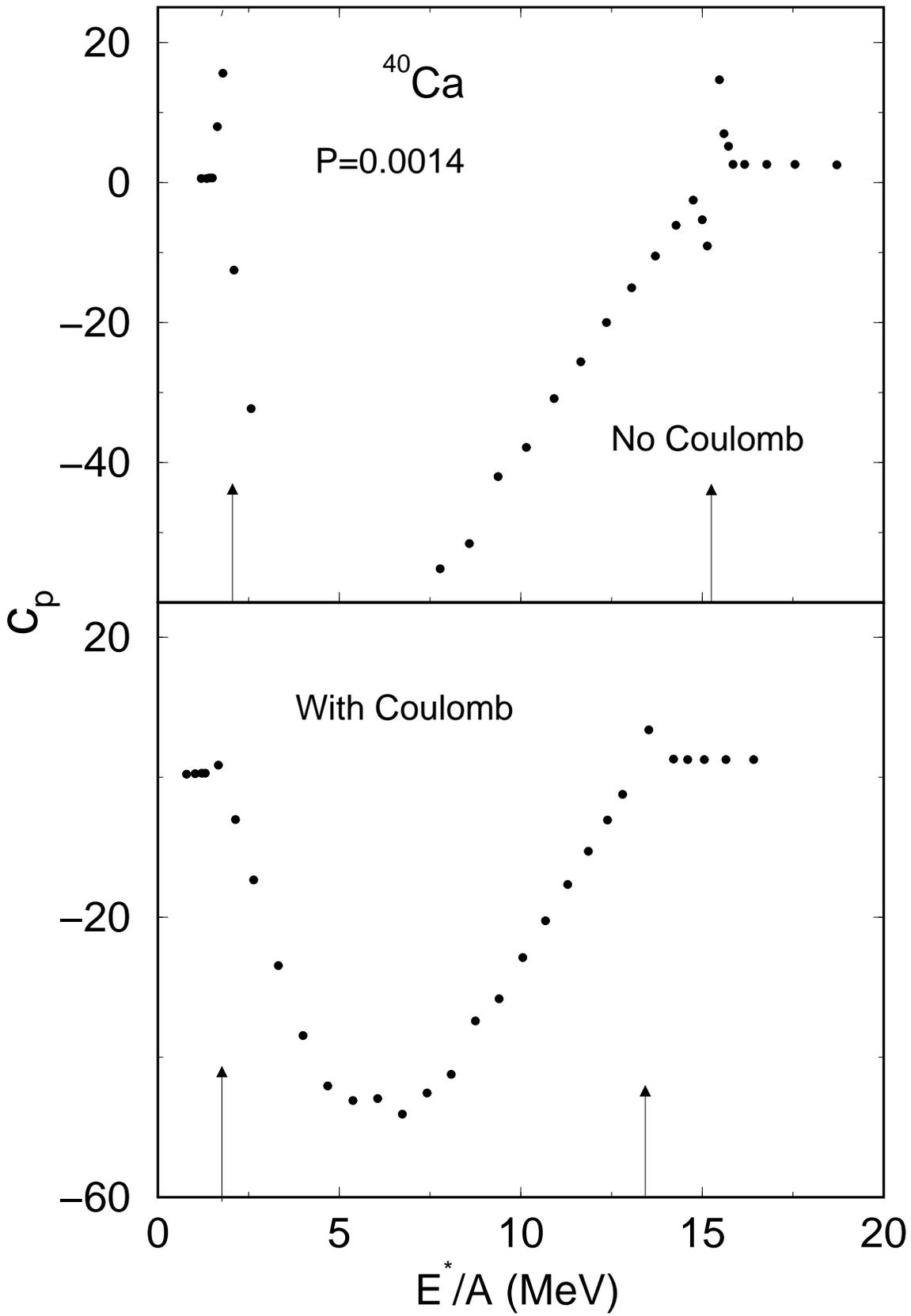
\item[Fig.\ 7] Heat capacity at constant pressure for the system $^{40}$Ca
with (bottom panel) and without (top panel) Coulomb interaction.
The arrows indicate the points of discontinuity.
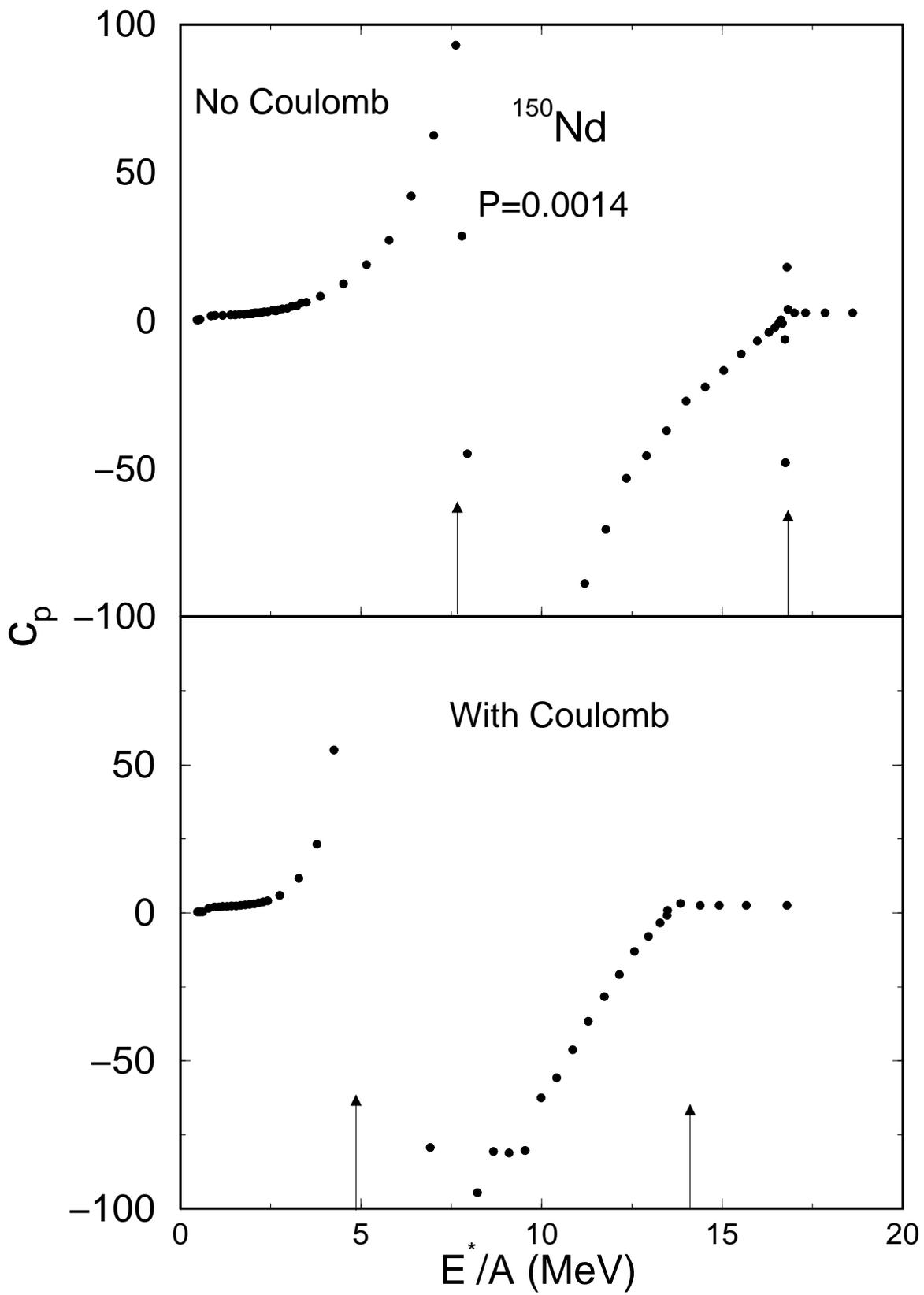
\item[Fig.\ 8] Same as Fig. 7 for the system $^{150}$Nd. 
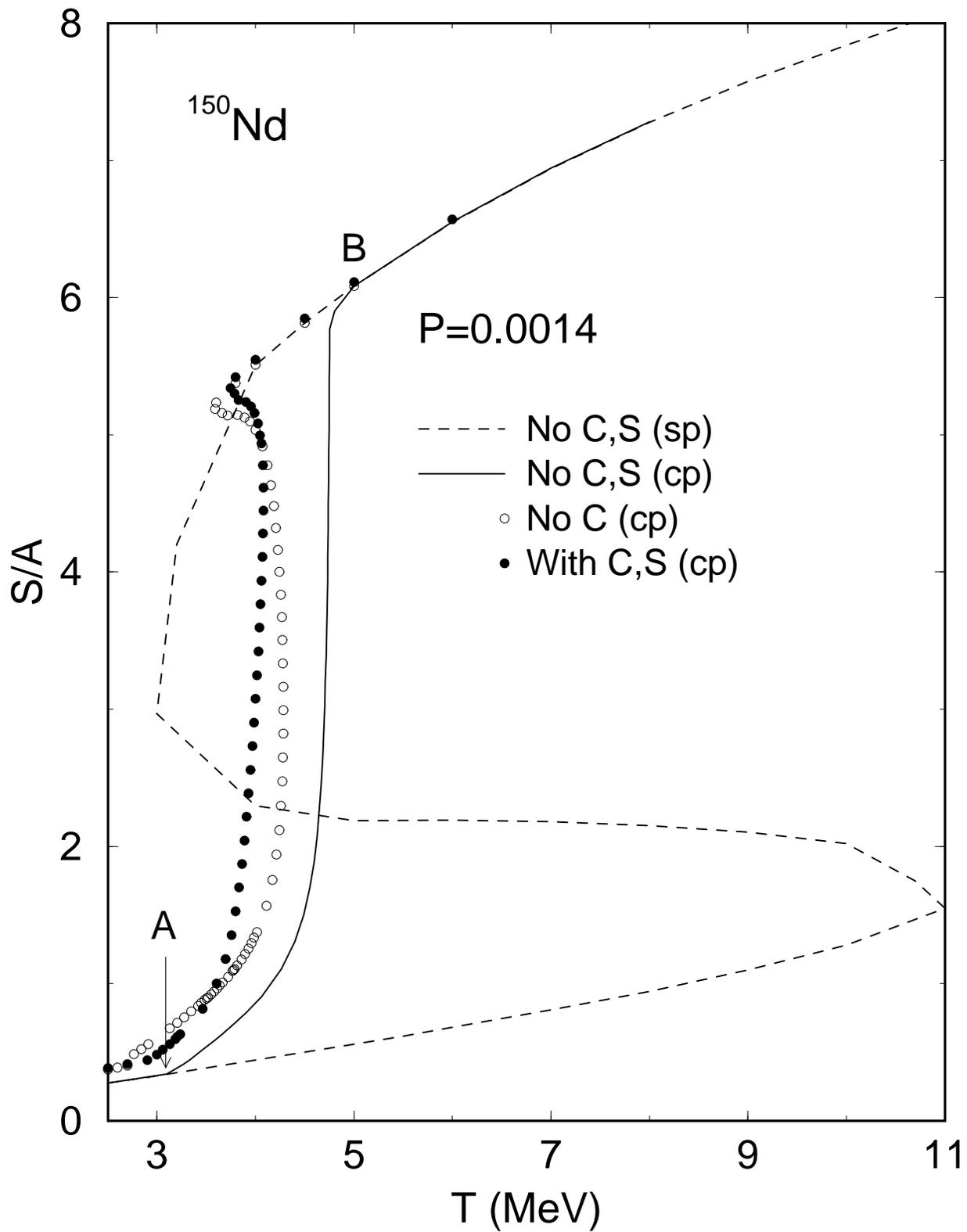
\item[Fig.\ 9] Entropy per particle $S/A$ as a function of temperature
for the system $^{150}$Nd at $P$=0.0014 MeV fm$^{-3}$. The notations
are the same as in Fig. 6.
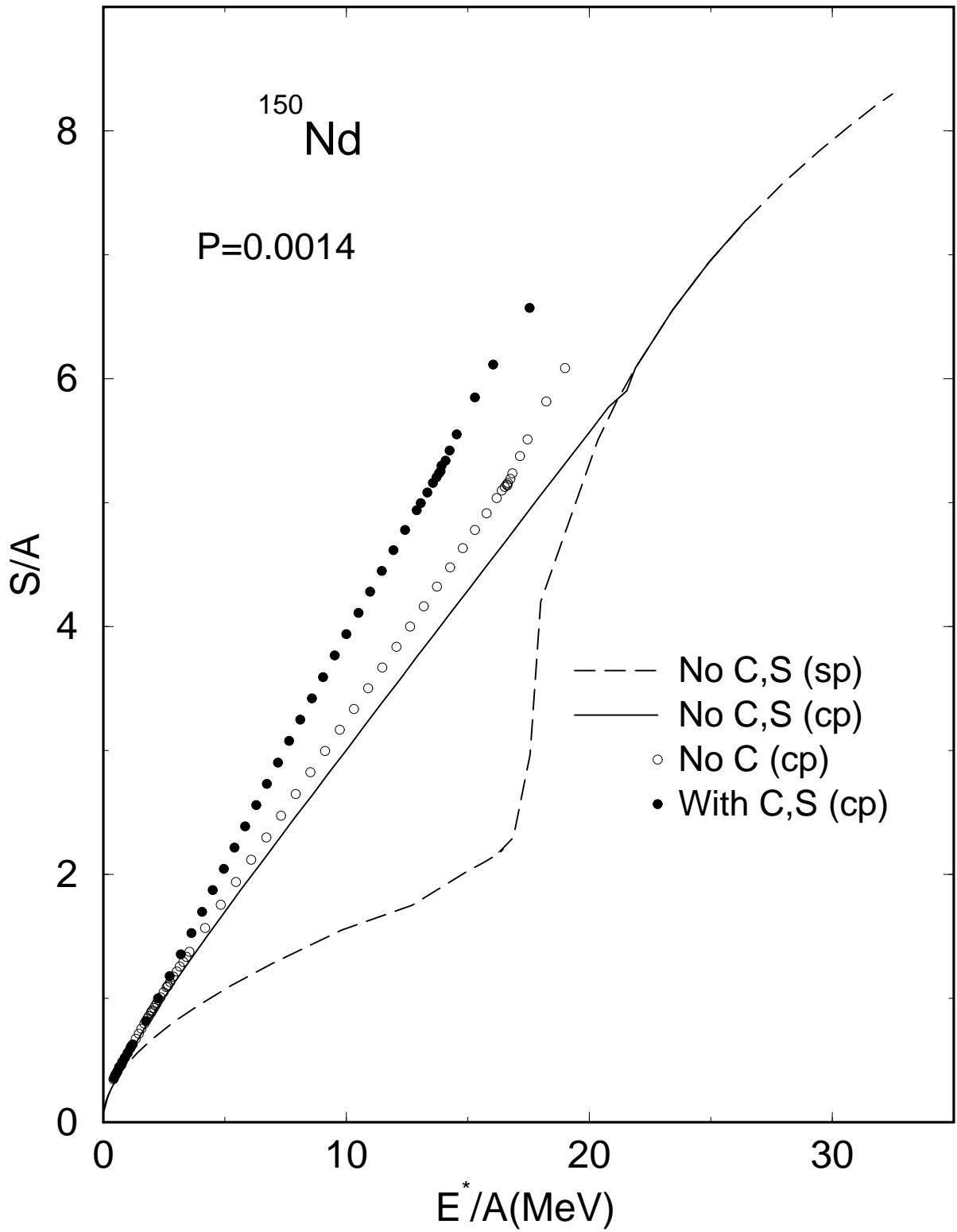
\item[Fig.\ 10] Entropy per particle as a function of excitation energy
per particle $E^*/A$ for the case same as in Fig. 9. The notations
are as in Fig. 6.
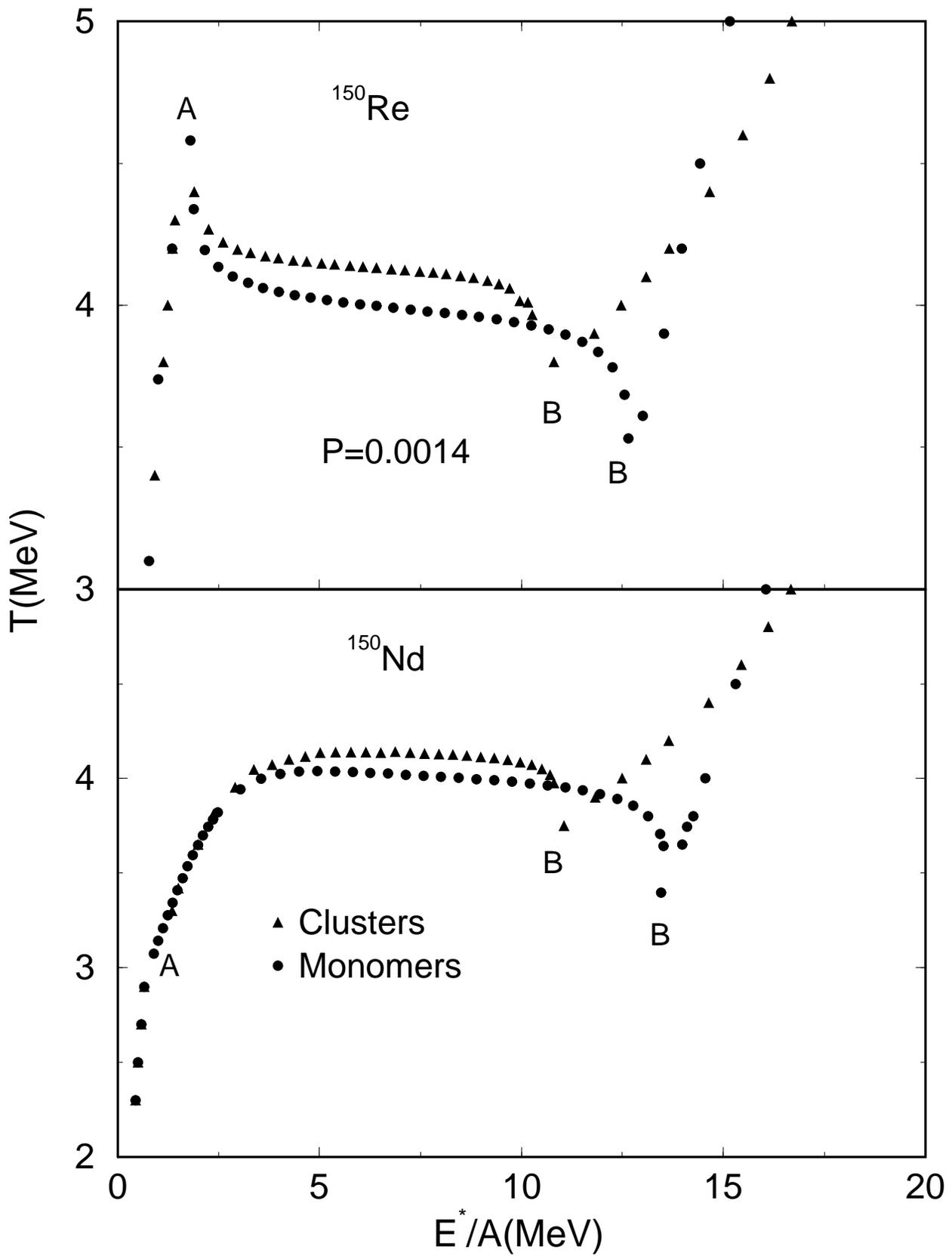
\item[Fig.\ 11] Caloric curve with monomers (filled circles) 
and clusters (filled triangles) in the vapor phase
for the systems $^{150}$Re (top panel) and $^{150}$Nd
(bottom panel).
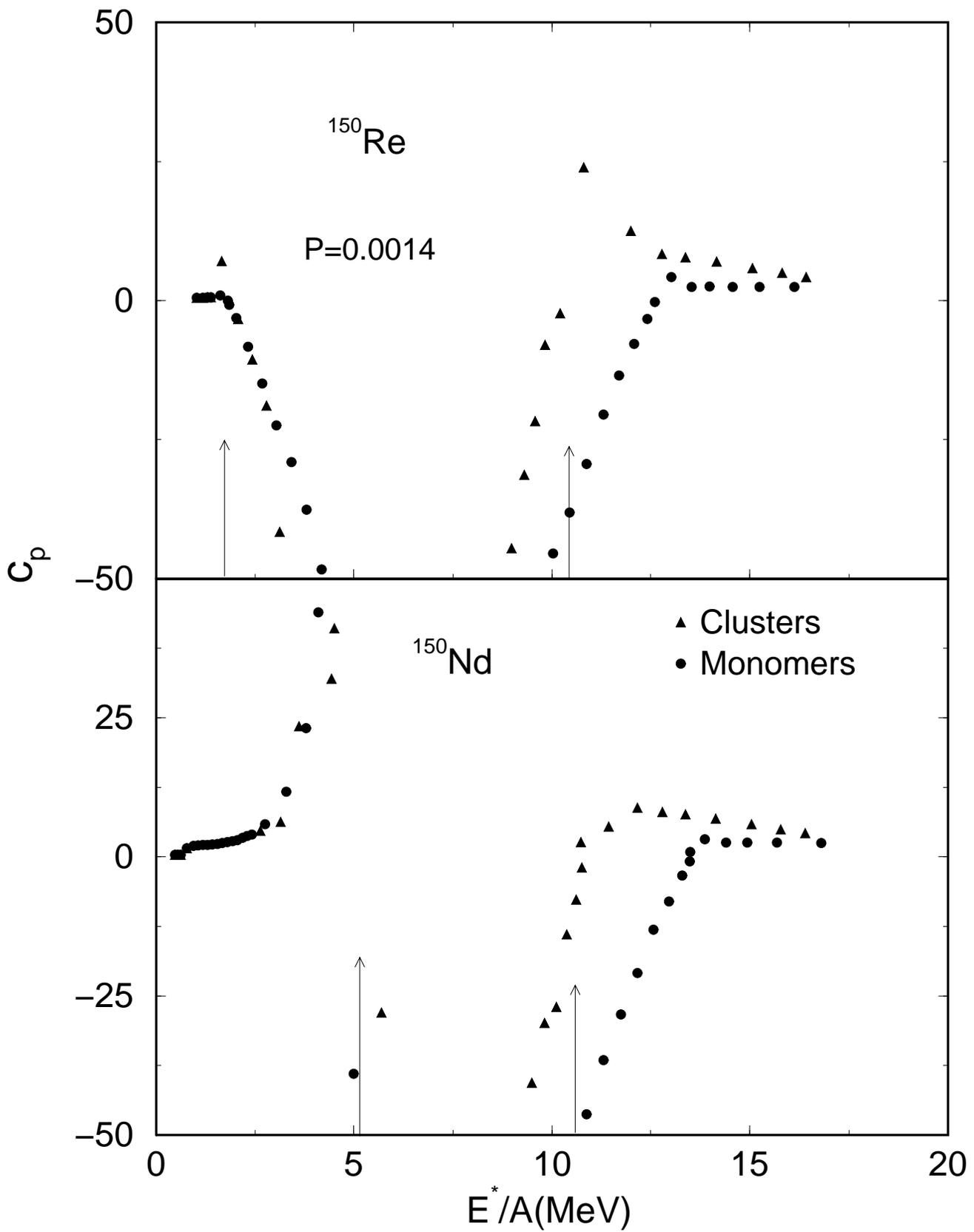
\item[Fig.\ 12] Heat capacity at constant pressure for the system 
$^{150}$Re (top panel) and $^{150}$Nd (bottom panel). 
 The arrows indicate the points of discontinuity.
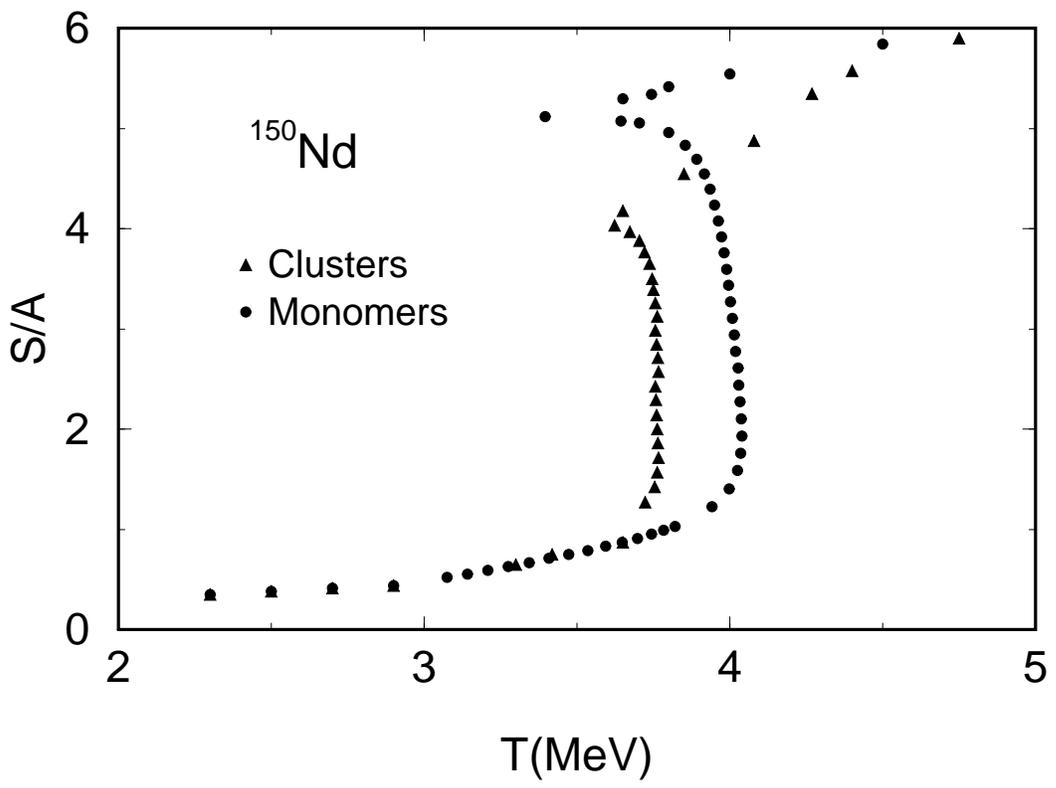
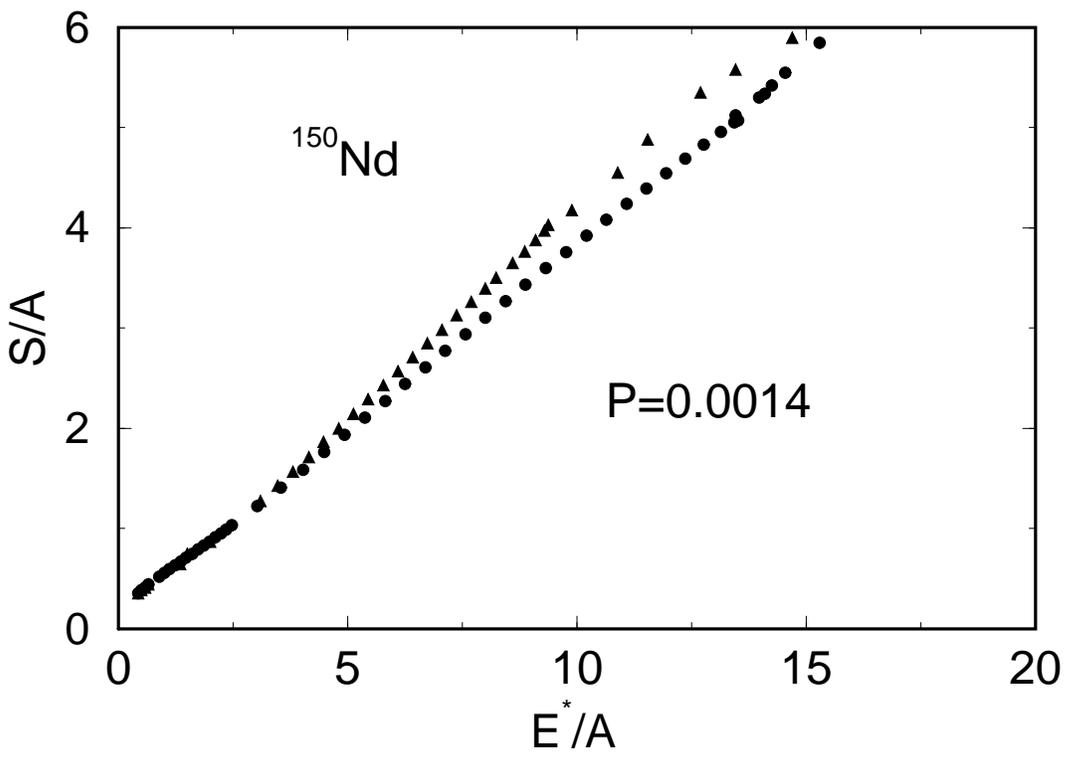
\item[Fig.\ 13] Entropy as a function of temperature (top panel)
and excitation energy (bottom panel) for the system $^{150}$Nd at
P=0.0014 MeV fm$^{-3}$.
\end{itemize}
\end{document}